\documentclass[aps,prl,superscriptaddress,twocolumn]{revtex4-1}

\usepackage{amsmath}
\usepackage{amssymb}
\usepackage{graphicx}
\usepackage{color}
\usepackage[colorlinks,citecolor=blue,linkcolor=blue,urlcolor=blue]{hyperref}

\newcommand{\bra}[1]{\langle #1|}
\newcommand{\dd}{\mathrm{d}}

\newcommand{\ket}[1]{|#1\rangle}
\newcommand{\mbb}[1]{\mathbb{#1}}

\newcommand{\mcal}[1]{\mathcal{#1}}
\newcommand{\mr}[1]{\mathrm{#1}}

\begin{document}
\title{Spectroscopic Feature of Quantum Many-Body Scar States}

\author{Wenjie Wei}
\affiliation{Kavli Institute for Theoretical Sciences and CAS Center for Excellence in Topological Quantum Computation, University of Chinese Academy of Sciences, Beijing 100190, China}

\author{Long Zhang}
\email{longzhang@ucas.ac.cn}
\affiliation{Kavli Institute for Theoretical Sciences and CAS Center for Excellence in Topological Quantum Computation, University of Chinese Academy of Sciences, Beijing 100190, China}
\affiliation{Hefei National Laboratory, Hefei 230088, China}

\date{\today}

\begin{abstract}
We study the dynamical correlations of nonintegrable systems with quantum many-body scar (QMBS) states generated by a ladder operator. The spectral function of the ladder operator has an exact $\delta$-function peak induced by the QMBS states. As a concrete example, we show that in the one-dimensional (1D) spin-$1$ Affleck-Kennedy-Lieb-Tasaki model, the spectral function of two-magnon excitations exhibits a characteristic bow-tie shape composed of a $\delta$-function resonance peak at momentum $k=\pi$ and a continuum spectrum elsewhere. The two-magnon excitations can be observed with the resonant inelastic X-ray scattering spectroscopy on quasi-1D nickelates and other spin-$1$ antiferromagnetic materials, thus our work paves the way to detecting the (approximate) QMBS states in realistic materials.
\end{abstract}
\maketitle 

{\bf Introduction.}	Quantum many-body scar (QMBS) states are atypical nonthermalizing eigenstates in certain nonintegrable quantum systems \cite{Turner2018, Turner2018a, Moudgalya2018, Moudgalya2018a, Ho2019, Khemani2019, Lin2019, Choi2019, Serbyn2021, Papic2022, Moudgalya2022, Chandran2023}. In a generic nonintegrable quantum many-body system, it has been postulated that the physical observables of highly excited eigenstates vary smoothly with the energy up to small fluctuations that vanish in the thermodynamic limit, thereby an eigenstate is physically indistinguishable from the microcanonical ensemble. This eigenstate thermalization hypothesis (ETH) lays the conceptual foundation of quantum statistical physics \cite{Deutsch1991, Srednicki1994a, Rigol2008a, DAlessio2016a, Deutsch2018}. However, certain nonintegrable quantum systems have QMBS eigenstates, whose physical properties strongly deviate from the adjacent eigenstates in the energy spectrum and the microcanonical ensemble. In particular, their entanglement entropy exhibits subvolume-law scaling instead of the volume-law scaling expected for the generic eigenstates obeying the ETH \cite{Turner2018, Turner2018a, Moudgalya2018a}. Therefore, the QMBS states violate the ETH, and thus have attracted intense research interest recently.

A variety of QMBS states can be constructed with the spectrum generating algebra (SGA) formalism \cite{Moudgalya2018, Mark2020}. Starting from a reference eigenstate $\ket{G}$ with energy $\epsilon_{G}$, a tower of QMBS states $\ket{S_{n}}$ ($n\in\mbb{N}$) are obtained by iteratively applying a ladder operator $Q^{\dagger}$ to $\ket{G}$, $\ket{S_{n}}=(Q^{\dagger})^{n}\ket{G}$. Here, the ladder operator $Q^{\dagger}$ is a summation over local (finite-range) operators and satisfies $[H,Q^{\dagger}]\ket{S_{n}}=\xi Q^{\dagger}\ket{S_{n}}$ ($n\in\mbb{N}$). Therefore, these $\ket{S_{n}}$ form an equally spaced tower of QMBS states, $H\ket{S_{n}}=(\epsilon_{G}+n\xi)\ket{S_{n}}$. The SGA formalism applies to the one-dimensional (1D) Affleck-Kennedy-Lieb-Tasaki (AKLT) model \cite{Moudgalya2018}, the generalized spin-$1$ XY model \cite{Schecter2019}, the generalized Hubbard model \cite{Mark2020a}, etc. It has been generalized to the quasisymmetry group formalism of QMBS states \cite{Ren2021}.

The tower of QMBS states leads to the constant revival of physical quantities in the real-time evolution if the initial state has a substantial overlap with the QMBS states \cite{Turner2018, Turner2018a}; In contrast, physical quantities of a generic initial state obeying the ETH quickly relax to the thermal equilibrium state. Such anomalous long-term revival phenomena due to the QMBS states have been observed in quantum simulation experiments with Rydberg atom arrays and other platforms \cite{Bernien2017, Bluvstein2021, Su2023a, Liang2024}.

In this work, we study the dynamical correlations of quantum many-body systems with scar states. We show that the spectral function of the SGA ladder operator has an exact $\delta$-function peak due to the QMBS states. In particular, we find that in the 1D AKLT model, the spectral function of two-magnon excitations exhibits a characteristic bow-tie shape, which consists of a $\delta$-function resonance peak at momentum $k=\pi$ and continuum spectra at other generic momenta. The resonance peak persists but broadens if the Hamiltonian is slightly deformed away from the AKLT model. We propose that the two-magnon spectrum can be observed with the resonant inelastic X-ray scattering (RIXS) experiments on quasi-1D nickelates and other spin-$1$ antiferromagnetic (AF) materials, thereby providing a spectroscopic evidence of (approximate) QMBS states in realistic materials.

{\bf SGA and spectral function in AKLT model.} The 1D AKLT model is the spin-$1$ AF Heisenberg model deformed by a biquadratic term \cite{Affleck1987a, Affleck1988a},
\begin{equation}
H=\sum_{l=1}^{L}P_{l,l+1}=\sum_{l=1}^{L}\Big(\frac{1}{3}+\frac{1}{2}\vec{S}_{l}\cdot\vec{S}_{l+1}+\frac{1}{6}\big(\vec{S}_{l}\cdot\vec{S}_{l+1}\big)^{2}\Big),
\end{equation}
in which $P_{l,l+1}$ is the projection operator into the total spin-$2$ subspace of the two sites $l$ and $l+1$. The ground state of the AKLT model with the periodic boundary condition is the unique common eigenstate of all $P_{l,l+1}$ with eigenvalue $0$. It can be exactly represented by the following matrix-product state (MPS) \cite{Klumper1993, Schollwock2011a, Orus2014},
\begin{equation}
\ket{G}=\sum_{\{s_{l}\}}\mr{Tr}\Big(\prod_{l=1}^{L}M^{[s_{l}]}\Big)\ket{s_{1}s_{2}\cdots s_{L}},
\end{equation}
in which $\ket{s_{l}}$ ($s_{l}=0,\pm1$) are the eigenstates of $S_{l}^{z}$ with eigenvalue $s_{l}$. The matrices are given by
\begin{equation}
M^{[1]}=
\begin{pmatrix}
0 & \sqrt{2} \\
0 & 0
\end{pmatrix},
M^{[0]}=
\begin{pmatrix}
-1 & 0 \\
0 & 1
\end{pmatrix},
M^{[-1]}=
\begin{pmatrix}
0 & 0 \\
-\sqrt{2} & 0
\end{pmatrix}.
\end{equation}
The ground state is a short-range correlated disordered state with the spin correlation length $\xi=1/\ln 3$ \cite{Affleck1987a, Affleck1988a}. The AKLT model has a finite excitation gap \cite{Affleck1987a, Affleck1988a}, which is in accord with the Haldane conjecture on integer-spin AF Heisenberg chains \cite{Haldane1983a, Haldane1983}. Besides, the AKLT model is a prominent example of symmetry-protected topological order with degenerate edge states in the open boundary condition \cite{Gu2009, Pollmann2010}.

A tower of QMBS states of the AKLT model can be constructed by applying the following ladder operator
\begin{equation}
Q_{\pi}^{\dagger}=\frac{1}{\sqrt{L}}\sum_{l=1}^{L}(-1)^{l}(S_{l}^{+})^{2}
\end{equation}
to the ground state \cite{Moudgalya2018}, $\ket{S_{n}}=(Q_{\pi}^{\dagger})^{n}\ket{G}$ ($0\leq n\leq L/2$). They satisfy $[H,Q_{\pi}^{\dagger}]\ket{S_{n}}=2Q_{\pi}^{\dagger}\ket{S_{n}}$, thus the $\ket{S_{n}}$ form a tower of eigenstates with energy $\epsilon_{n}=2n$. These states can also be efficiently represented by MPS with virtual bond dimensions bounded by $L$, thus their entanglement entropy exhibits the subvolume-law scaling \cite{Moudgalya2018a}, which is a hallmark of QMBS states. Besides, $Q_{\pi}=(Q_{\pi}^{\dagger})^{\dagger}$ generates another tower of QMBS states $\ket{\tilde{S}_{n}}=Q_{\pi}^{n}\ket{G}$ with energy $\tilde{\epsilon}_{n}=2n$, because $Q_{\pi}$ and $Q_{\pi}^{\dagger}$ are related by the spin rotation symmetry transformation $R_{x}(\pi)=\otimes_{l}e^{i\pi S_{l}^{x}}$, $Q_{\pi}=R_{x}(\pi)Q_{\pi}^{\dagger}R_{x}(\pi)^{\dagger}$.

Moreover, the operator $Q_{k}^{\dagger}$ satisfies
\begin{equation}
Q_{k}^{\dagger}=\frac{1}{\sqrt{L}}\sum_{l=1}^{L}e^{ilk}(S_{l}^{+})^{2}=\frac{1}{\sqrt{L}}\sum_{q}S_{q}^{+}S_{k-q}^{+},
\end{equation}
in which $S_{q}^{+}=L^{-1/2}\sum_{l=1}^{L}e^{iql}S_{l}^{+}$. $Q_{k}^{\dagger}$ creates a two-magnon excitation with total momentum $k$,  thus the QMBS state $\ket{S_{1}}=Q_{\pi}^{\dagger}\ket{G}$ is an exact bound state of two-magnon excitations. However, $Q_{k}^{\dagger}\ket{G}$ ($k\neq \pi$) is not an exact eigenstate in general.

In this work, we shall consider the retarded Green's function of two-magnon excitations at zero temperature,
\begin{equation}
G(\omega,k)=-i\int_{0}^{\infty}\dd t\,e^{i(\omega+i0^{+})t}\bra{G}[Q_{k}(t),Q_{k}^{\dagger}(0)]\ket{G}.
\end{equation}
The spectral function $A(\omega,k)=-\frac{1}{\pi}\mr{Im}G(\omega,k)$ can be expressed in the Lehmann's spectral representation,
\begin{equation}
\begin{split}
A(\omega,k) &= \sum_{\alpha}|\bra{\alpha}Q_{k}^{\dagger}\ket{G}|^{2}\delta(\omega-\epsilon_{\alpha}) \\
&-\sum_{\alpha}|\bra{\alpha}Q_{k}\ket{G}|^{2}\delta(\omega+\epsilon_{\alpha}),
\end{split}
\end{equation}
in which the summation is taken over the complete eigenstate basis $\ket{\alpha}$ with the excitation energy $\epsilon_{\alpha}\geq 0$. Moreover, we have $A(\omega,k)=-A(-\omega,k)$, because $Q_{k}^{\dagger}$ and $Q_{k}$ are related by the symmetry transformation composed of the spin rotation $R_{x}(\pi)$ and the space inversion $\mcal{I}$. For the AKLT model, the spectral function at momentum $k=\pi$ has $\delta$-function resonance peaks due to the QMBS states $\ket{S_{1}}=Q_{\pi}^{\dagger}\ket{G}$ and $\ket{\tilde{S}_{1}}=Q_{\pi}\ket{G}$, $A(\omega,k=\pi)\propto \delta(\omega-2)-\delta(\omega+2)$. For $k\neq \pi$, $A(\omega,k)$ exhibits a continuum spectrum, thus forming a characteristic bow-tie shape as we shall see in the following numerical results.

We have the following sum rule for the AKLT model,
\begin{equation}
\int_{0}^{\infty}\dd \omega\,A(\omega,k)=\frac{4}{3},\quad\forall k.
\end{equation}
This is proved by noting that the integral is given by the static structure factor of the two-magnon operator at the ground state,
\begin{equation}
\bra{G}Q_{k}Q_{k}^{\dagger}\ket{G}
=\frac{1}{L}\sum_{ll'}e^{ik(l'-l)}\bra{G}(S_{l}^{-})^{2}(S_{l'}^{+})^{2}\ket{G},
\end{equation}
and the correlation function of the two-spin-flip operator satisfies
\begin{equation}
\bra{G}(S_{l}^{-})^{2}(S_{l'}^{+})^{2}\ket{G}=\frac{4}{3}\delta_{ll'},\quad L\rightarrow \infty.
\end{equation}
The vanishing of the correlation function for $l\neq l'$ is derived from the strict diluted AF order of the AKLT ground state \cite{Affleck1987a, Affleck1988a, DenNijs1989}.

\begin{figure*}[tp]
\centering
\includegraphics[width=\linewidth]{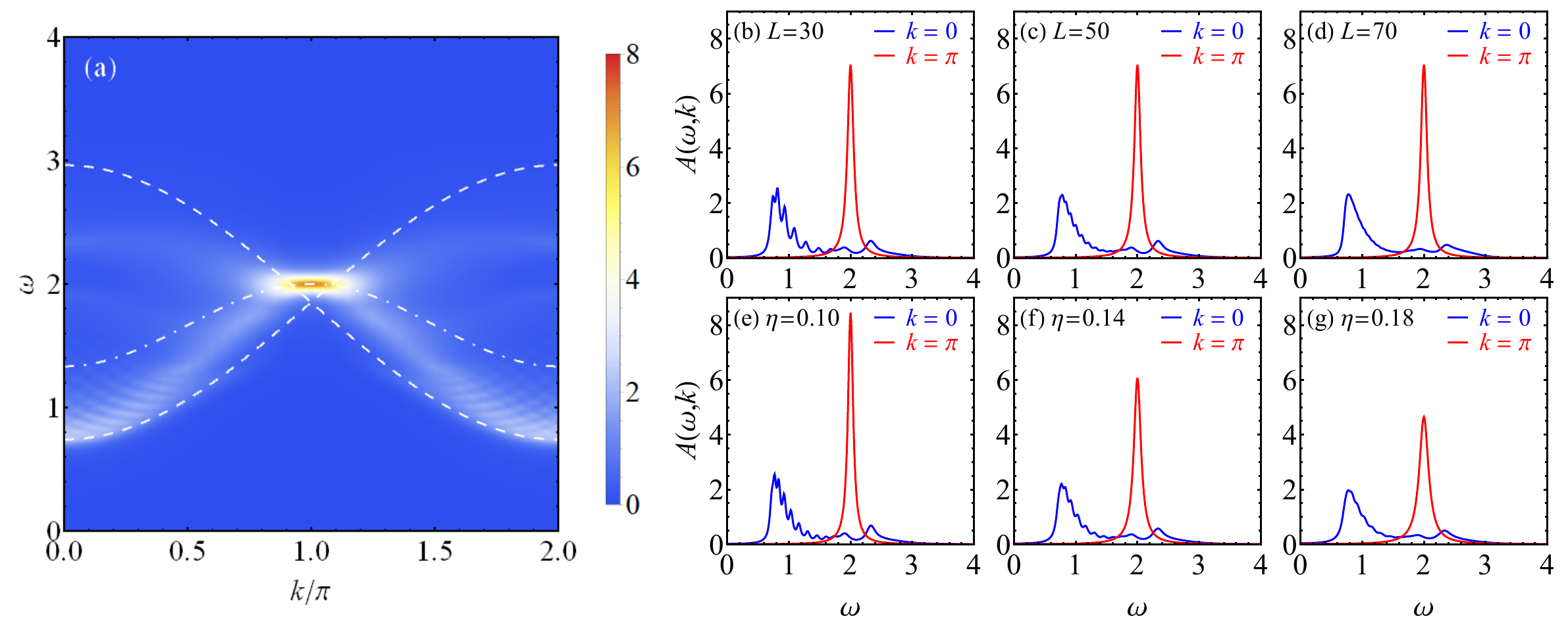}
\caption{(a) The two-magnon spectral function of the AKLT model at the ground state. The dotdashed curve is the average energy $\varepsilon_{k}$ given in Eq. (\ref{eq:average}). The dashed curves are the approximate upper and lower edges in Eq. (\ref{eq:edges}) obtained with the single-mode approximation of one-magnon excitations. Numerical results are obtained on a lattice with $L=40$ and periodic boundary condition. The order of the truncated series of the KPM $N=1000$. The effective Lorentz broadening width $\eta=0.12$. (b--d) Comparison of $A(\omega,k)$ at $k=0$ and $\pi$ for different lattice sizes $L=30$, $50$, and $70$ with the same Lorentz broadening width $\eta=0.12$. The sharp resonance peak at $k=\pi$ is barely changed, while the discrete peaks in $A(\omega,k)$ at $k=0$ merge into a continuum spectrum in the thermodynamic limit. (e--g) Comparison of $A(\omega,k)$ at $k=0$ and $\pi$ with different Lorentz broadening widths $\eta=0.10$, $0.14$, and $0.18$ for the same lattice size $L=40$. The sharpness of the peak at $k=\pi$ is only limited by the Lorentz broadening, which is consistent with a $\delta$-function peak induced by the QMBS states.}
\label{fig:main}
\end{figure*}

{\bf Numerical results.} The two-magnon spectral functions of the AKLT model and the deformed Hamiltonian in Eq. (\ref{eq:deform}) are calculated with the MPS-based density-matrix renormalization group (DMRG) \cite{White1992, White1993, Schollwock2005, Schollwock2011a, Orus2014} and the kernel polynomial method (KPM) \cite{Weisse2006a}. The ground state $\ket{G}$ in the periodic boundary condition is obtained with the DMRG algorithm and represented by an MPS. In the KPM, the Hamiltonian is first rescaled such that all eigenvalues are bounded in the interval $[-1,1]$, then its spectral function $A(\omega,k)$ for $\omega>0$ is approximated by a truncated series \cite{Weisse2006a},
\begin{equation}
A(\omega,k)=\frac{1}{\pi\sqrt{1-\omega^{2}}}\Big(g_{0}\mu _{0}+2\sum_{n=1}^{N-1}g_{n}\mu _{n}T_{n}(\omega)\Big)
\label{eq:expansion}
\end{equation}
of the Chebyshev polynomials of the first kind $T_{n}(\omega)=\cos(n\arccos \omega)$, in which the order of truncation $N$ is set large enough to guarantee convergence of the series. The expansion coefficients
\begin{equation}
\mu _{n}=\bra{G}Q_{k}T_{n}(H)Q_{k}^{\dagger}\ket{G}\equiv \langle\alpha_{0}\ket{\alpha_{n}},
\end{equation}
in which $\ket{\alpha_{n}}=T_{n}(H)Q_{k}^{\dagger}\ket{G}$ are calculated by applying $Q_{k}^{\dagger}$ and $H$ (represented by matrix-product operators) to $\ket{G}$ according to the iteration relation, $\ket{\alpha_{0}}=Q_{k}^{\dagger}\ket{G}$, $\ket{\alpha_{1}}=H\ket{\alpha_{0}}$, and $\ket{\alpha_{n+1}}=2H\ket{\alpha_{n}}-\ket{\alpha_{n-1}}$ ($n\geq 1$). In Eq. (\ref{eq:expansion}), the Lorentz kernel factors $g_{n}=\sinh(\lambda(1-n/N))/\sinh\lambda$ are introduced to improve the behavior of the truncated series. This is equivalent to convolving the spectral function with the Lorentz lineshape function with width $\eta=\lambda/N$ \cite{Weisse2006a}, thus it effectively mimics the finite resolution of experimental apparatus. Afterwards, the frequency in the spectral function is rescaled back to the energy range of the original Hamiltonian.

The two-magnon spectral function of the AKLT model is calculated with the KPM and plotted in Fig. \ref{fig:main}. A prominent feature is the resonance peak at $k=\pi$. As illustrated in panels (b--g), the sharpness of this resonance peak is only limited by the Lorentz broadening width, thus it is consistent with the $\delta$-function peak induced by the QMBS states as proved before. In contrast, the spectral function at $k\neq \pi$ approaches a continuum in the thermodynamic limit, indicating that the two-magnon excitations do not form coherent quasiparticles. Therefore, the spectral function $A(\omega,k)$ has a characteristic bow-tie shape composed of the $\delta$-function resonance peak induced by the QMBS states at $k=\pi$ and a continuum spectrum elsewhere.

The average energy of the two-magnon excitation $Q_{k}^{\dagger}\ket{G}$ can be calculated analytically with the method in Ref. \cite{Affleck1988a}. The result is given by
\begin{equation}
\varepsilon_{k}=\frac{\bra{G}Q_{k}HQ_{k}^{\dagger}\ket{G}}{\bra{G}Q_{k}Q_{k}^{\dagger}\ket{G}}=\frac{1}{3}(5-\cos k),
\label{eq:average}
\end{equation}
and is plotted in Fig. \ref{fig:main} (a). The energy of the two-magnon excitations with total momentum $k$ can also be approximated by summing over the energy of two one-magnon excitations, $E_{k}=\xi_{q}+\xi_{k-q}$, in which the one-magnon energy $\xi_{q}$ is estimated with the single-mode approximation \cite{Arovas1989},
\begin{equation}
\xi_{q}\simeq \frac{\bra{G}S_{-q}^{-}HS_{q}^{+}\ket{G}}{\bra{G}S_{-q}^{-}S_{q}^{+}\ket{G}}=\frac{5}{27}(5+3\cos q),
\end{equation}
thus we have $E_{k}\simeq \frac{10}{27}\big(5+3\cos(k/2)\cos(k/2-q)\big)$, which is bounded by
\begin{equation}
\frac{10}{27}\big(5-3|\cos(k/2)|\big)\lesssim E_{k}\lesssim \frac{10}{27}\big(5+3|\cos(k/2)|\big).
\label{eq:edges}
\end{equation}
This turns out to be a good approximation of the upper and lower edges of the two-magnon spectrum, which is illustrated in Fig. \ref{fig:main} (a).

\begin{figure}[tp]
\centering
\includegraphics[width=\columnwidth]{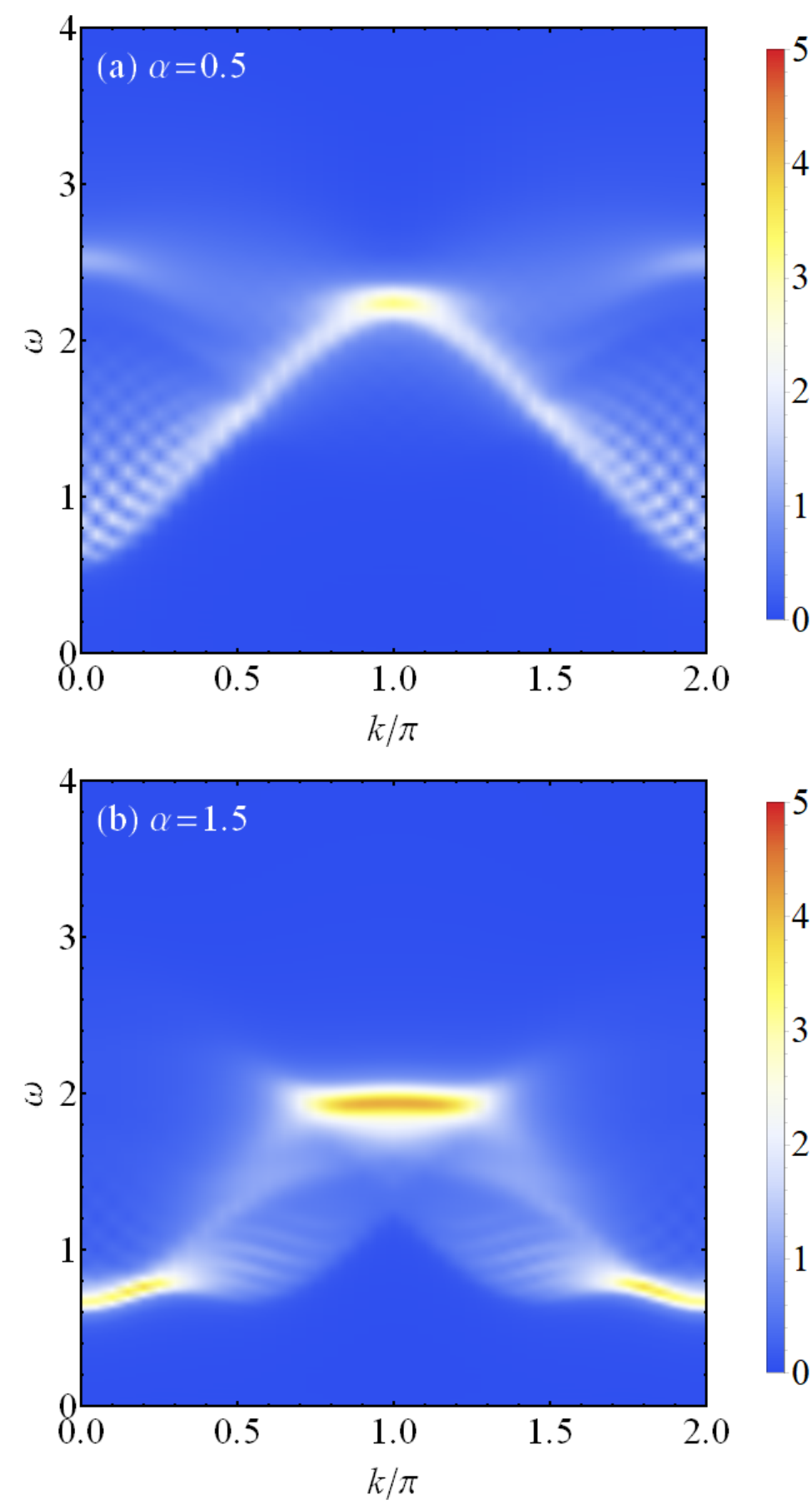}
\caption{Two-magnon spectral function $A(\omega,k)$ of the deformed Hamiltonian in Eq. (\ref{eq:deform}) with (a) $\alpha=0.5$ and (b) $\alpha=1.5$. The lattice size $L=40$. The order of the truncated series of the KPM $N=1000$. The effective Lorentz broadening width $\eta=0.22$.}
\label{fig:deform}
\end{figure}	

We then consider the following spin-$1$ bilinear-biquadratic Hamiltonian \cite{Bursill1995},
\begin{equation}
H=\sum_{l=1}^{L}\Big(\frac{1}{3}+\frac{1}{2}\vec{S}_{l}\cdot\vec{S}_{l+1}+\frac{\alpha}{6}\big(\vec{S}_{l}\cdot\vec{S}_{l+1}\big)^{2}\Big),
\label{eq:deform}
\end{equation}
which is a deformation of the AKLT model, and is relevant for realistic quasi-1D AF materials, e.g. LiVGe$_{2}$O$_{6}$ \cite{Millet1999, Lou2000}. Its two-magnon spectral functions $A(\omega,k)$ for $\alpha=0.5$ and $1.5$ are plotted in Fig. \ref{fig:deform}. The exact SGA relation breaks down in the deformed Hamiltonian and $Q_{\pi}^{\dagger}\ket{G}$ is not an exact eigenstate, thus the resonance peak at $k=\pi$ is not a $\delta$-function any more; Nonetheless, the spectral function maintains the overall bow-tie shape with a sharp resonance peak (though with finite width) at $k=\pi$ and a broad continuum elsewhere. This is reminiscent of the atypical eigenstates with low entanglement entropy found in the deformed Hamiltonian in Ref. \cite{Moudgalya2018a}.

{\bf Proposal for RIXS experiments.} We propose that the characteristic bow-tie shape of the two-magnon spectral function induced by the QMBS states can be detected with the RIXS spectroscopy on quasi-1D spin-$1$ AF nickelates, e.g., CsNiCl$_{3}$ and Ni(C$_{2}$H$_{8}$N$_{2}$)$_{2}$NO$_{2}$ClO$_{4}$ (NENP). The one-magnon excitations of these materials were observed by the inelastic neutron scattering spectroscopy and exhibit an energy gap of spin excitations \cite{Buyers1986, Tun1990, Ma1992, Regnault1994}, which supports the Haldane gap of spin-$1$ AF chains. In the RIXS of spin-$1$ magnetic materials, the differential scattering cross section is given by the spectral function of an effective scattering operator $R^{\varepsilon_{i}\varepsilon_{o}}_{j}$ \cite{Haverkort2010}. When the polarization of the incoming and the outgoing X-ray beams are given by $\varepsilon_{i}=\varepsilon_{o}^{*}=(1,i)/\sqrt{2}$, i.e., they are left and right circularly polarized, respectively, the spin transfer in the scattering process is $\Delta S=+2$, then the scattering operator at site $j$ is given by $R_{j}^{\varepsilon_{i}\varepsilon_{o}}=\sigma^{(0)}+\sigma^{(2)}(S_{j}^{+})^{2}$, in which $\sigma^{(0)}$ and $\sigma^{(2)}$ are constant factors \cite{Haverkort2010}. Therefore, the RIXS precisely measures the two-magnon spectral function $A(\omega,k)$ of the spin-$1$ magnetic materials, and is able to detect the characteristic bow-tie shape induced by the QMBS states in quasi-1D AF materials.

{\bf Summary.} We have shown that the QMBS states constructed with the SGA formalism produce a $\delta$-function resonance peak in the spectral function of the SGA ladder operator. As a concrete example, we show that the two-magnon spectral function in the AKLT model exhibits a characteristic bow-tie shape composed of a resonance peak at $k=\pi$ induced by the QMBS states and a continuum spectrum elsewhere. This overall bow-tie shape persists in the deformed Hamiltonian with a broadened resonance peak. We propose that the two-magnon spectral function can be observed in the RIXS experiments on quasi-1D nickelates and other spin-$1$ AF materials, thereby providing a spectroscopic evidence of (approximate) QMBS states in realistic materials. Our results of the resonance peak in the spectral function applies to other QMBS systems generated with the SGA formalism as well, e.g., the spin-$1$ generalized XY models in higher dimensions. Therefore, similar spectroscopic feature is also expected in these systems.

{\bf Note added.} Upon the completion of this work, we became aware of a recent work \cite{Ren2024}, in which the spectral function of QMBS systems is studied from the perspective of the quasisymmetry group and the quasi-Nambu-Goldstone mode.

\acknowledgments
L.Z. is grateful Xue-Rong Liu and Xingye Lu for helpful discussions. Part of the numerical simulations was carried out with the ITensor package \cite{Fishman2022}. This work is supported by the National Natural Science Foundation of China (No. 12174387), the Chinese Academy of Sciences (YSBR-057 and JZHKYPT-2021-08), and the Innovative Program for Quantum Science and Technology (No. 2021ZD0302600).

\bibliography{../../BibTex/library}
\end{document}